\documentclass[10pt,letterpaper,twocolumn]{article} %% two column, final layout

\usepackage{ol}
\usepackage{hyperref}
\usepackage{amsmath}

\begin{document}

\twocolumn[
\title{Optical Coherence Spectro-Tomography by all-Optical Depth-Wavelength analysis}

\author{L. Froehly, M. Ouadour, L. Furfaro, P. Sandoz, T. Gharbi}

\address{D\'epartement d'Optique PM Duffieux, Institut
FEMTO-ST, UMR CNRS $6174$, Universit\'e de Franche-Comt\'e, $25030$
Besan\c{c}on Cedex, France}

\author{P. Leproux, G. Huss, V. Couderc}

\address{XLIM - UMR CNRS 6172, 123,avenue Albert Thomas, 87060 Limoges Cedex, France}

% Do not use \email or \homepage here. E-mail and URL can be given just before references.
\begin{abstract}
Current spectroscopic optical coherence tomography (OCT) methods
rely on a posteriori numerical calculation. We present an
alternative for accessing optically the spectroscopic information in
OCT, i.e. without any post-processing, by using a grating based
correlation and a wavelength demultiplexing system. Conventional
A-scan and spectrally resolved A-scan are directly recorded on the
image sensor. Furthermore, due to the grating based system, no
correlation scan is necessary. In the frame of this paper we present
the principle of the system as well as first experimental results.
\end{abstract}

]

For a decade the interest for Optical Coherence Tomography (OCT) has
been growing in the field of biomedical imaging. Main reasons are
the non destructive character of these methods, the image resolution
down to the micrometer scale either in-depth or in-plane and the
capability to perform optically \emph{in vivo} non-destructive biopsies.
Tomographic images can be obtained by different OCT configurations
which are classified in two main families: Time-Domain OCT (TD-OCT)
and Fourier Domain OCT (FD-OCT)\cite{Fercher2003}.

For a few years a new trend is to complement the reconstruction of in-depth 3D tissue structure by functional information as a help in medical diagnosis.
In this way solutions that supply indications on the actual biological
metabolism of the inspected tissues were reported; for instance:
polarization OCT imaging\cite{Hee1992}, spectroscopic
OCT\cite{Watanabe2000} or CARS-OCT\cite{Vinegoni2003}. Our work associates also a functional signature to in-depth OCT
microstructure reconstructions. We propose an all-optical device for
the spectro-tomographic characterization of the inspected tissues.
We access optically to a spectro-tomogram characterizing the
depth-wavelength behavior of a sample line ($x_{0},y_{0},z$). This
spectro-tomogram corresponds to an usual A-scan that is
spectrally-resolved on a continuous set of spectral bands. The width of the latter is determined by the setting of the experimental set-up. Furthermore this
spectro-tomogram is obtained without any scanning since the depth
exploration is performed by a temporal correlator based on a static
diffraction grating.

\begin{figure}[htb]
\centerline{\includegraphics[width=8cm]{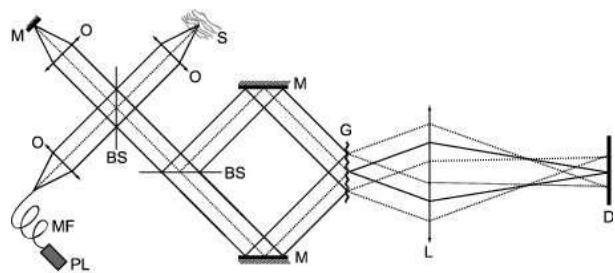}} \caption{
Experimental set-up (top-view): PL pump laser; MF microstructured
fiber; O microscope objectives; BS beam splitter; S sample; G
diffraction grating; L spherical lens; D
detector.}\label{set_up_top_view}
\end{figure}

\begin{figure}[htb]
\centerline{\includegraphics[width=6cm]{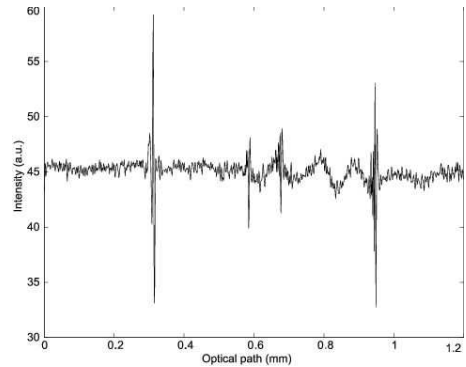}}
\caption{Usual time-domain A-scan as obtained experimentally without
scanning on an eosine solution layered between two microscope
coverslips.}\label{TD_A_scan}
\end{figure}

The experimental set-up used is composed of three main parts as
depicted in Fig. \ref{set_up_top_view}. Firstly the sample
information is encoded via a Linnik interferometer. The latter is
illuminated with a supercontinuum of light issued from a
microstructured optical fiber pumped by a Q-switched Nd-YAG
laser\cite{Tombelaine2005}. Since this low coherence light source has a transversally singlemode emission, the sample is illuminated only
along the axial point spread function of the microscope objective
used (therefore the OCT information finally obtained with
depth-wavelength resolution corresponds also to this sample volume).
Secondly the light beams issued from the Linnik interferometer are
directed toward a second interferometer. In this Mach-Zehnder-like
interferometer the output beamsplitter is replaced by a transmission
diffraction grating disposed in the perpendicular direction. Because
of the incident angles of the two beams, the transverse direction of
the diffraction grating introduces a time-delay $\tau$ varying
linearly between the recombined beams. The sample depth is thus
encoded across the grating that forms a time correlation axis. The
inherent principle of that kind of temporal correlator was first
introduced in $1957$\cite{Connes1957} for spectroscopy. In 1991 it
has been applied to intermodal dispersion measurements in optical
fibers\cite{Brun1992}. More recently some adaptations of this set-up
were proposed for optical
tomography\cite{Verrier1997,Zeylikovich1997,BenHoucine2004,Froehly2006}. A key
property of that configuration is to reduce significantly the
carrier frequency of the interference fringes. This effect can be
explained by the change of the average propagating direction of the
interfering wavefronts after diffraction in the $-1$ order.  The
third part of the set-up is a simple imaging system that forms the
image of the diffraction grating on a two-dimensional CCD image
sensor. Then the lines of the CCD camera encode the depth of the
sample and a A-scan is displayed without scanning along the image
lines (since all the light incident on the diffraction grating is
issued from the same sample line, the different lines of the
recorded image carry the same depth information). The cascade of the
two interferometers should result in autocorrelation. In practise,
interfering beams are cross-polarized in both interferometers
thanks to polarization multiplexing (quarter and half waveplates
are not represented in Fig. \ref{set_up_top_view} for the sake of
clarity). Thus we perform intercorrelation instead of
autocorrelation and the detected signal can be expressed as:

\begin{eqnarray}
C(x)=I+2\mathcal{R}e[\int_{\nu}R(\nu)S(\nu)e^{-j2\pi(\frac{2x}{\gamma
c}sin\theta_{i})\nu}e^{j\frac{4\pi x}{\gamma \Lambda}}d\nu]
%C(x)=I_{0}+2\mathcal{R}e\int_{\Delta\lambda}E(\nu)S^{*}(\nu)e^{j2\pi(\frac{2x}{c}sin\theta_{i})\nu}e^{-j\frac{4\pi}{\Lambda}}d\nu
\end{eqnarray}
 where x is the horizontal coordinate on the CCD camera lines, I the background
intensity, $\mathcal{R}e$ designs the real part, $R(\nu)$ and
$S(\nu)$ are the spectral distributions of the reference and sample
beams respectively, $\theta_{i}$ is the incidence angle on the
diffraction grating, $\Lambda$ is the grating period and $\gamma$ is
the imaging system magnification. The grating effect appears in the
term $e^{j\frac{4\pi x}{\gamma \Lambda}}$ that introduces the fringe
frequency change. (The reference beam $R(\nu)$ is assumed to be real
while $S(\nu)$ can be either real or complex depending on the
optical sample properties). Fig. \ref{TD_A_scan} presents the
intensity distribution along an image line as recorded for the
inspection of a homemade sample. The latter is made of a 2\% Gifrer
Eosine solution layered by capillarity between two microscope
coverslips. The four interfaces (air-glass; glass-solution;
solution-glass, glass-air) are clearly visible and the thickness of
both the coverslip and the eosine solution can be retrieved from
this intensity distribution after depth calibration. Those data
could be used for spectroscopic analysis as obtained classically by
numerical windowed Fourier Transform. The light source spectrum
extends from $350nm$ to $1750nm$. In this experiment an effective
bandwidth of $100nm$ centered around $550nm$ is chosen and selected
through the size of the imaging lens. This bandwidth should results in a
full depth resolution of about $6\mu m$. In practice,mainly due to sample dispersion, it is clear on Fig. \ref{TD_A_scan} that the resolution is lower ($10-15\mu m$).

At this stage of the set-up description, the device constitutes a
Linear TD-OCT system in a configuration different from anterior
works\cite{Verrier1997,Zeylikovich1997,BenHoucine2004,Hauger2003} .

\begin{figure}[htb]
\centerline{\includegraphics[width=8cm]{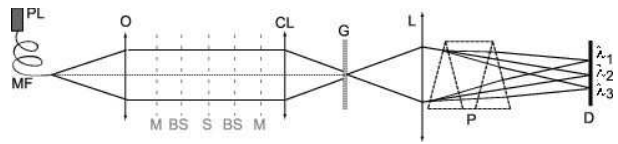}}
\caption{Experimental set-up (side-view): PL pump laser; MF
microstructured fiber; O microscope objectives; BS beam splitter; S
sample; G diffraction grating; L spherical lens; CL cylindrical
lens; P prism; D detector.}\label{set_up_side_view}
\end{figure}

\begin{figure}[htb]
\centerline{\includegraphics[width=7cm]{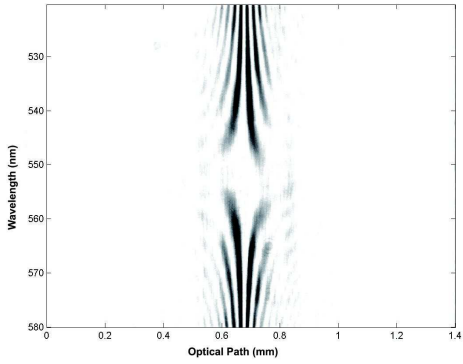}}
\caption{Experimental spectro-tomographic signal for
autocorrelation (mirror as sample).}\label{Trace_autocorr}
\end{figure}

Complementary elements are necessary for the obtention of a
spectro-tomogram. These optical elements affect the light
propagation only along the vertical direction and appear in Fig.
\ref{set_up_side_view}, that presents a side view of the set-up. The
key-element is the prism that changes the output imaging system into
a spectroscope. Then the vertical direction of the CCD camera
becomes a spectral axis. The spectroscope resolution is tuned
through the position of the cylindrical lenses inserted in the
Mach-Zehnder arms and that focus the light beams incident on the
grating in a horizontal line, whose height controls the spectral
resolution. In this complete system configuration each image line is
associated to a particular wavelength and is illuminated by a
restricted bandwidth. Therefore each image line provides a
spectrally-resolved A-scan and the whole image forms a
spectro-tomogram. Fig. \ref{Trace_autocorr} presents the recorded
spectro-tomograph as a mirror is used as sample. In that case we
obtain a signal related to the light source autocorrelation since
the sample beam is not modified. The spectral resolution is $1nm$
corresponding to a depth resolution of $300\mu m$ (given by the
relation $\Delta\nu.\Delta t=1$). In this result each line
corresponds to the autocorrelation of a $1nm$ bandwidth centered
around the corresponding wavelength. The progressive variation of
the fringe period is due to the wavelength dependence of the
diffracted angle. For $\lambda=550nm$, the interfering beams are
collinear after diffraction and no fringes are visible. The
spectro-tomographic information can be expressed as:
\begin{eqnarray}
C(x,y)=I
+2\mathcal{R}e[\int_{\nu}F(y,\nu)R(\nu)S(\nu)e^{-j2\pi(\frac{2x}{\gamma
c}sin\theta_{i})\nu}\nonumber
\\ e^{j\frac{4\pi x}{\gamma \Lambda}}d\nu]
%C(x,y)=I_{0}+2\mathcal{R}e\int_{\Delta\lambda}F(y,\nu)E(\nu)S^{*}(\nu)e^{j2\pi(\frac{2x}{c}sin\theta_{i})\nu}e^{-j\frac{4\pi}{\Lambda}}d\nu
\end{eqnarray}
where y is the vertical coordinate on the CCD camera and $F(y,\nu)$
is the spectroscope response for the line y.

\begin{figure}[htb]
\centerline{\includegraphics[width=7cm]{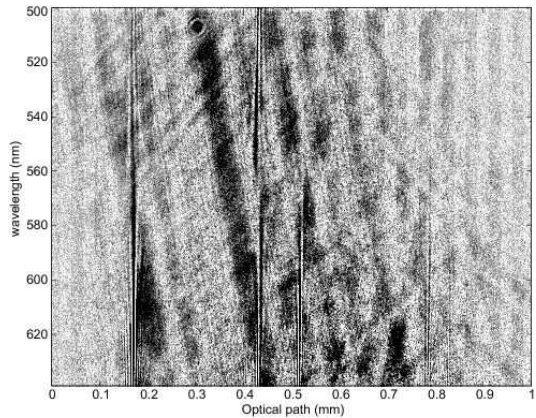}}
\caption{Instantaneous spectro-tomographic signal with
depth-resolved spectral absorption of the eosine
layer.}\label{eosinespectro}
\end{figure}

The inspection of the eosine sample resulted in Fig.
\ref{eosinespectro}. In this case the spectral resolution was fixed
to $12nm$, leading to a depth resolution of $25\mu m$. We observe
that the four sample interfaces are visible only in the lower part
of the figure, i.e. outside the absorption band of
eosine\cite{these1996}. Wavelengths corresponding to the upper part
of the figure are absorbed by the solution and no light returns from
the last interfaces that are no more detectable. This spectral
information is obtained optically and instantaneously. It
complements the reconstruction of the in-depth microstructure of the
sample. This result demonstrates clearly the depth-wavelength
capabilities of the proposed method for optical coherence
spectro-tomography. The lateral resolution is determined by the
objective used(5X, N.A.=0.1)and was measured to be $20\mu m$ (USAF pattern).

At this stage of system development, signal to noise ratio (SNR)
measurements would not be significant of the ultimate method
capabilities since detection elements are not optimized yet. However
one may notice already the following points. Our detection scheme
differs from TD-OCT one's while reconstructed A-scans are quite
similar. In one hand no scanning is required and the integration
time on the detector can be much longer. On the other hand dynamics
and noise performances of image sensors are usually worse than those
of photodiodes or PMTs. Finally our detection is closer to FD-OCT
one's since each A-scan information is contained on one CCD line.

A clear interest of an all optical processing system appears for the
spectroscopic analysis of OCT responses. In that case, the spectral
resolution and the spectral bandwidth of interest are actually
determined by the optical components of the device. This point is an
advantage over post-processing techniques such as windowed Fourier
transforms which are constrained by digital sampling and
discretization parameters. Furthermore spectro-tomograms obtained
optically are available instantaneously and that point can be
important for applications in which a high measurement rate is
required. The main drawback of our grating based correlation system
is the DC part which decreases the dynamic range available
for signal detection. This problem could be solved using a
particular kind of CMOS detection already used in parallel
TD-OCT\cite{Laubscher2002}. This detector allows both heterodyne
detection and DC filtering in real-time. The next step of our work
will be to demonstrate the feasibility of the implementation of this
kind of detection with our system and the influence of this on the
system performances (SNR, sensitivity and dynamic range available
for signal detection).

Acknowledgements: We acknowledge the French
ANR for funding this work (ANR-05-JCJC-0187-01).

\end{document}